# An End-to-End Encryption Solution for Enterprise Content Applications

Chaoting Xuan {cxuan} @gatech.edu

**Abstract**

The content host services (like Dropbox, OneDrive, and Google Drive) used by enterprise customers are deployed either on premise or in cloud. Because users may store business-sensitive data (contents) in these hosting services, they may want to protect their data from disclosure to anyone else, even IT administrators. Unfortunately, even contents (files) are encrypted in the hosting services, they sometimes are still accessible to IT administrators today. The sensitive data could be exposed to public if the IT administrator turns malicious (like disgruntled employee) or his account is compromised by hackers.

We propose an end-to-end encryption (E2EE) solution to address this challenge. The user data is encrypted at client side (mobile device) and remains encrypted in transit and at rest on server. Specifically, we design a new method to allow master secret recover and escrow, while protecting them from being accessed by malicious administrators. In addition, we present a content (file) encryption scheme that achieves privacy, and granular access control. And it can be seamlessly integrated with major content host services used by business users today.

## 1. Introduction

Nowadays, content hosting services used by enterprise customers are deployed either on premise or in cloud. IT administrators can access employees' contents (files) stored in hosting services and break the data privacy, even when the contents are encrypted. In a PKI based organization, IT administrators not only have the highest privileges in company's computer systems, but also own and manage the identity systems like active directory. This gives them the power to break employees' data security.

For example, an IT administrator can acquire the private keys of an employee at key generation, recovery or escrow phases [1][2]. Moreover, he sometimes can install spyware on employee's device and steal the sensitive data on client side. So, when an IT administrator turns malicious (like disgruntled employee) or his account is hacked by attackers, the sensitive contents of employees stored in hosting services are in danger of being compromised, no matter whether they are encrypted or not. In the paper, we call both malicious IT administrators and hackers who compromise the IT administrators' accounts *root attackers*. To best of our knowledge, ***how to protect employees' cloud data from root attackers is not thoroughly tackled in academic and industry.***

Currently, major cloud storage service vendors like Google, Microsoft and Dropbox don't provide E2EE services to customers because of business and legal considerations, e.g., they may want to scan customers' cloud data to provide extra services [3]. However, this is not desirable to many enterprise customers who require high-standard data security and privacy, especially for highly regulated industries such as all current healthcare. Other vendors try to solve this problem by adding E2EE on top of existing cloud storage services provided big vendors [4][5], or building the E2EE capabilities into their own cloud storage services [6][7]. Unfortunately, all these solutions only address the concerns of untrusted cloud storages, but not root attackers who can get the logon credentials and encryption keys of target employees, and decrypt their cloud contents.

To defending against root attackers, we leverage trusted execution environments (TEE) [8][9], and construct a new E2EE scheme, which can not only prevent root attackers from accessing employees' master secret used to encrypt the contents, but also offer the features friendly to enterprise customers, such privacy and granular access control. It's also a practical solution and can be integrated with any content applications with minor engineering efforts.

## 2. Background

### 2.1 Current E2EE Solutions

Today, a typical (and simplified) encryption process of a E2EE solution is as below: 1. a private and public key pair is generated on user's device; 2. The user uses his password to encrypt the private key via a password-based key derivation algorithm like PBKDF2[10] or Bcrypt [11]; 3. The encrypted private key and origin public key are send to a remote key server; 4. when a content application on device is going to encrypt a file, it checks if the private key is available locally, if not, the content application downloads the encrypted private key and ask user to type in the password to decrypt the private key; 4. A file key (symmetric key) is generated and used to encrypt the file , and the file key is encrypted using the private key; 5. The encrypted file and file key are sent to the file hosting server.

To share the file with another user, the content application locates the recipient's public key from key server, and uses it to encrypt the file key, which is sent to the hosting service along with the encrypted file and file key. The recipient can use his private key to decrypt the file key and file sequentially. Moreover, to assure the integrity and authenticity, a signing key pair could be created and used to sign the encrypted file and file key(s).

### 2.2 Technical Challenges

In this paper, we define the *master secret* as one encryption key that can be used to directly decrypt a target encrypted content and keys. In 2.1, master secret is the private key used to encrypt the file key. There are three ways to protect master secrets in practice: 1. Using password to derive shared keys and then encrypt the master secrets [12], as what 2.1 does; 2. Regenerating the master secrets via mnemonic phrases, which are written down to papers [13]; 3. Storing master secrets to smart cards or other external devices like USB drive [14].

However, all three methods have weaknesses: password could be forgotten; the papers recording mnemonic phrases and smart cards (USB drives) could be lost. If any of these situations occur, encrypted could be unrecoverable and lost forever [15]. The tougher challenges come from law enforcements like US government. As required by laws, some companies ask employees to send unencrypted master secrets to key escrow services [2], so that law enforcement can access the master secrets and decrypt the contents of target employee under certain circumstances. Unfortunately, there is no guarantee that these key escrow services (and stored master secretes) are inaccessible to root attackers.

In addition to master secret protection challenges, current E2EE-based content applications suffer one or more of following weaknesses [4][5][6][7] : 1. Not supporting file sharing; 2. Coarse-grained file access control; 3. Privacy information disclosure; 4. Susceptible to rollback attacks; 5. Not supporting link-based file sharing; 6. Not flexible to support various file hosting services. In this paper, we propose a new E2EE solution that is composed of two parts: *master secret protection* (in Section 3) and *content encryption* (in Section 4). The first part is designed to protect master secret from root attackers; the second part is to address the common encryption weaknesses suffered by current E2EE-based content applications.

### 2.3 Trusted Execution Environments

A Trusted Execution Environment (TEE) is a secure area inside a main processor. It runs in parallel of the operating system, in an isolated environment. It guarantees that the code and data loaded in the TEE are protected with respect to confidentiality and integrity. Trusted applications running in a TEE have access to the full power of a device's main processor and memory, whereas hardware isolation protects these components from user installed applications running in the main operating system. Software and cryptographic isolations inside the TEE protect the different contained trusted applications from each other. Secure Enclave and TrustZone are two popular TEE solutions used on mobile devices [8][9].

### 2.4 Scope and Threat Model

Assume there is a PKI system in a company, called IDENTITY, which provides full-fledged identity services to employees. When IDENTITY issues certificates (and key pairs) to an employee, it's assumed that the private key is generated on employee's local devices and are never disclosed to IDENTITY server in plaintext. An enterprise content application is composed of client component and server component, namely CLIENT and SERVER. SERVER can be deployed on premise, in cloud or both. Employees' company devices support TEEs, where their private keys are stored, and CLIENTs run.

Here, we assume that adversaries are root attackers who can take control of SERVER, company's PKI system and key escrow service. Device TEEs are resistant to root attackers, so employees' private keys and CLIENT's operations are secure. In addition, root attackers are able to create fake employee accounts with valid PKI and IDENTITY identities, and they can use these fake accounts to run CLIENTs on the devices under their control. Further, root attackers can masquerade some real employees by stealing their PKI private keys and passwords when their devices don't have TEE support. So, some CLIENTs could be malicious.

In the new E2EE solution, we aim to maintain the confidentiality and integrity of a target employee's contents stored in SERVER. The availability of encrypted contents is not guaranteed here, because root attackers can simply delete all the contents in SERVER to make them unavailable.

### 3. Master Secret Protection

Without loss of generality, we don't specify the exact cryptography algorithms like (RSA encryption) and their parameters (like 256-bit AES key) used in this paper, because we believe they should be application (and implementation) dependent and specific. Instead, we use the general words like "symmetric encryption" to represent these algorithms.

As mentioned above, an employee has PKI key pair (PK, SK) issued by IDENTITY. A master key $K\_master$ is generated for the content application in CLIENT. Note an employee may have multiple copies of CLIENTs run on multiple devices. All CLIENTs of this employee share same PKI key pair and master key. For each file, the employee uses CLIENT to generate file key $K\_file$.

Since file key (K_file) is encrypted by master key (K_master), so master secret of this employee is his master secret. The PKI private key (SK) is used to access other employees' contents shared to this employee, and when it is lost and unrecoverable, this employee can directly ask the owner employee to re-encrypt this content with new PKI public key (PK') that this employee holds. So, SK is not part of master secret. The details of content encryption are described in Section 4.

### 3.1 Master Secret Recovery

We employ two recovery methods: *password-protected server recovery* and *distance-bounded social recovery*. The first method is trivia: employee creates a password and use it to encrypt his master secret through a password-based key derivation algorithm. The encrypted master secret is stored to SERVER. When recovery is needed, the employee downloads encrypted master secret from SERVER to CLIENT, and then type in his password to decrypt the master secret. When employee forget his password, or the encrypted master secret is not available on SERVER, e.g., accidently deleted by owner, he need resort to distance-bounded social recovery.

#### 3.1.1 Distance-bounded Social Recovery

The basic idea of distance-bounded social recovery is to split the master secret into multiple sub-secrets (called *shards*) using some secret sharing algorithm (Shamir's Secret Sharing is a classic one [16]). And then distribute these shards to the CLIENTs of peer employees nearby. The physical distance control is achieved by running the secret sharing through peer-to-peer communication over short distance communication channels such Bluetooth Low Energy (BLE) or Near-field Communication (NFC). Before sending the shards, the owner employee is required to explicitly confirm the authenticities of recipient employees based on his witness of their physical existence.

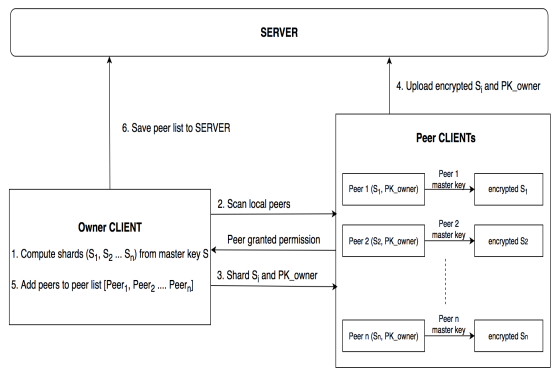

**Fig 1. Master Secret Sharing Process**

Secret sharing algorithms possess fault tolerance: as long as k of n recipients (n > k) are honest and benign, then original secret can be restored. Recall that some CLIENTs might be malicious, and this fault tolerance property is necessary. The communication between owner employee and peer employees are secured by using their PKI certificates, and protocol details are up to application. The secret sharing process for the master secret is as below:

1. Shard computation: using chose secret sharing algorithm to compute n shards, and output ($S_1$, $S_2$, … $S_n$)
2. Peer selection: owner CLIENT scans local peer-to-peer network; randomly chooses n peer CLIENTs running in TEEs and propose them to the owner by displaying peer employees' PIIs like names, emails and pictures (if possible). If owner cannot confirm the physical existence of a proposed peer employee, he should reject this proposal, otherwise, accept it.
3. For *i*th chosen peer CLIENT, Owner CLIENT sends it the *i*th shard ($S_i$) and public key (PK_owner).
4. *i*th chosen peer CLIENT confirms to owner CLIENT that ($S_i$, PK_owner) is received; then uses its master key to encrypt $S_i$ and upload the ciphertext of $S_i$ and PK_owner to SERVER.
5. Once getting the confirmation from *i*th peer CLIENT, owner CLIENT adds peer employee (PK_peer_i) to the peer list
6. Step 2 to 4 is repeated until n shards are successfully sent to n peer CLIENTs.
7. Owner CLIENT uploads the peer list to SERVER.

To reconstruct the master secret, owner CLIENT downloads peer list from SERVER, and visit each peer CLIENT to acquire shard. When k authenticate shards are accumulated, owner CLIENT computes the master secret from the k shards. Note a peer CLIENT should verify the public key of owner (PK) to make sure that it downloads, decrypts and sends the shard to right owner. When owner PKI key pairs are changed like certificate being expired, the owner should update its public key in peer employee's shard record with the new one.

To obtain the master secret, root attackers need compromise at least k peer CLIENTs in the peer list. One possible attack is that root attackers create k+ fake peer CLIENTs with k+ valid PKI key pairs; and exploit the peer selection to add the fake peers to owner's peer list. However, short-distance peer-to-peer communication and owner's visual verification significantly mitigate this attack scheme, yielding it almost impractical.

### 3.2 Master Secret Escrow

According threat model in 2.4, root attacker can compromise key escrow service (ESCROW), so it's unsafe for owner employee to directly place the plaintext of his master secret to ESCROW. One solution is to take advantage of distance-bounded social recovery and make peer CLIENTs to upload the encrypted shards to ESCROW (in step 4 of the secret sharing process mentioned above). On release of owner's master secret, law enforcements ask the peer employees to decrypt and turn in the shards in out-of-band means, e.g., using QR codes to pass the decrypted shards to law enforcements' devices; then the law enforcements can restore the master secret by running same secret sharing algorithm. To make this solution effective, two assumptions need hold: 1. Root attackers cannot successfully launch social engineering attacks and deceive the peer employees to submit the plaintexts of shards; 2. Owner employee need to be honest to distribute the authenticate shards to peer CLIENTs. The second assumption is not stronger than the case in which owner employee is required to directly submit plaintext of master key to ESCROW.

## 4. Content Encryption

Other than PKI key pair (PK, SK) and master key K_master, an employee generates two symmetric keys for each file: file encryption key (FEK) and file signature key (FSK). The FEK and FSK are used to differentiate between read and write access. Possession of only the FEK gives read only access to the file while possession of both the FEK and FSK allows read and write access. For example, a user with only the FEK cannot create a valid file because he cannot produce a valid file signature. In this way, a fine-grained file access control is achieved.

In addition, we require PKI key pair (PK, SK) to be elliptic-curve public-private key pair and serve Curve25519 cryptography algorithm [17]. Thus, any two employees automatically and implicitly share one symmetric key via elliptic curve Diffie-Hellman (ECDH) algorithm [18]. K_share_user1 represents one such key that an employee shares with user1. The employee uses K_share_user1 to share file with user1.

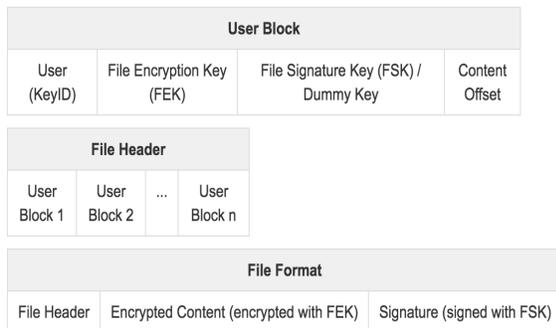

Fig 2. Encrypted File Format

### 4.1 File Format

To support various content hosting services, we don't directly modify the files' metadata, which sometimes are stored separately with file data in hosting services and requires different APIs to make changes. Instead, we concatenate all encryption metadata to file data, which is transparent to hosting services. This design significantly reduces the engineering efforts required to integrate this new E2EE scheme to any existing content hosting application.

An encrypted file is composed three parts: file header (encryption metadata), encrypted content and signature. Encrypted content represents the encrypted file data by using FEK, and signature represents the message authentication code (MAC) generated by using FSK to sign encrypted content. The file header is composed of group of user blocks. One user block contains the metadata that a user can use to read and/or write file data. Specifically, it contains the user ID (public key ID), FEK, FSK (or Dummy Key) and content offset.

When a file owner wants to share his file with another user (user1), he creates one user block as [user1 ID, FEK, X_KEY, content offset]. If write permission is granted, X_KEY is FSK; otherwise, it's a dummy key with same key size as FSK. Then, the user block is encrypted with K_share_user1, which is the implicit key shared between owner and user1. The first user block in a file header always belongs to file owner. Total number user blocks in a file header varies and depends on how many users that the owner grants file access to. When user1 accesses the owner's file, starting from the second user block, he sequentially decrypts rest user blocks until the one that contains his user ID. Then, he uses content offset to determine the starting position of encrypted content and decrypt the content using FEK. The encrypted file formation is shown in figure 2.

### 4.2 Freshness Guarantees

Freshness guarantees are required in order to prevent rollback attacks. A rollback attack involves misleading users into accessing stale data. For example, suppose Bob revokes Alice's permission to write to a file named foo. Alice does a rollback attack by replacing the new file header with an older version that she saved. The older version of the md-file has a valid signature and will hence verify correctly. Alice has now successfully restored her own write permissions to the file. Checking the file header for freshness would stop such an attack.

We use a hash tree [19] to guarantee freshness. There is a file header freshness file (*fhf-file*) located in every directory of a user's file system. This file contains the root of a hash tree built from all the *fhf-file*s in the directory and its subdirectories. For example, the *fhf-file* at the root of a user's home directory contains the root of the hash tree constructed from the user's *fhf-file* in the directory and *fhf-file*s under immediate subdirectories. A user's CLIENT will periodically time stamp the root *fhf-file* and sign it using his master key K_master. The update interval can be set by the user. When a malicious user replaces the new file header with an old one. The file owner's CLIENT will detect an unexpected root hash change and generate alert.

### 4.3 Identityless Privacy

File metadata (like access permissions) and file accessing records both can disclose privacy information [20]. For example, if root attackers find that a patient shares his health record file with a cancer doctor; or the cancer doctor visits his health record file multiple times, without decrypting the health record, they can conclude that the patient gets cancer with high probability. To address this issue, our E2EE scheme attempts to achieve the *identityless privacy*, which means that root attackers are blinded on who accesses a target file including owner himself.

In 4.1, all user blocks are encrypted by symmetric keys shared between owner and other users, and all user blocks have equal length (recall dummy key), so root attackers cannot base file header to derive any information regarding authorized users and their access permissions for a target file. However, a user's visits to SERVER and target file can disclose the privacy information and be tracked by root attackers. To solve this problem, a user can hide his network trace by using anonymous peer-to-peer network like Tor [21]; in addition, cloud storage services like Google Drive and Dropbox offer anonymous file sharing via links (anyone can access a file through randomized link as an anonymous user), and we can leverage this feature to hide the user's login to SERVER. Concretely, file owner generates a file link and shares it with a user; and the user access the file by traveling an anonymous peer-to-peer network and login to SERVER as an anonymous user.

## 5. Implementation

We implemented a PoC (proof of concept) of master secret sharing and recovery in Android platform. Bluetooth is used to find peers around to share and recover master secret. This PoC uses Android API BluetoothAdapter to find peers and uses *BluetoothServerSocket* and *BluetoothSocket* to implement the communications between user and peer CLIENTs. It can be extended to support other protocols like NFC and peer-to-peer WiFi etc. When peer-to-peer WiFi is used, Android framework provides *WiFiP2pManager* API to create a WiFi peer-to-peer network which doesn't require internet access. The implementation of Shamir's Secret Sharing is provided by *com.codahale.shamir* module [22] which is an open source module under Apache license. Shamir's Secret Sharing needs to work on finite field and in this module it uses GF(256). When owner tries to recruit peers, owner can decide the number of peers (n) and a threshold (k), where n is in [k, 255] and k is in [1, n].

The original *com.codahale.shamir* code will crash if n and k values are not legal, which is a bug. The PoC fixes it by checking the input n and k values, and prompting a message to ask user choose the right values. Inside the PoC, after user has made the decision, it will create a *Scheme* instance by using a secure random value and given n and k, then use the *scheme* instance to split the master secret. The output is a *HashMap* instance with entry <integer, byte array_of_byte>. Then each entry (key-value pair) will be sent to the selected peers. When user wants to recover the key, he needs to connect to those who have the valid key-value pairs, and once k key-value pairs have been collected, he can use the *Scheme* instance's *join()* method to reconstruct the master secret. When switching between string and byte array, we use standard charset UTF-8.

## 6. Conclusion

In conclusion, we propose a new E2EE solution that can defend against root attackers. It contains a new key recovery method called distance-bounded social recovery, based on which a new key escrow scheme is designed to prevent root attackers from accessing plaintexts of master secrets. In addition, a new content encryption scheme is devised to enjoy many plausible properties in comparison to current E2EE solutions, e.g., granular access control and identityless privacy.